\documentclass[preprint,showpacs,preprintnumbers,amsmath,amssymb]{revtex4}

\usepackage{amssymb}
\usepackage{psfig}
\usepackage{graphicx}% Include figure files

\newcommand{\rar}{\rightarrow}

\begin{document}

\preprint{Preprint M\'exico ICN-UNAM 05-04}

\title{A helium-hydrogenic molecular atmosphere of neutron star 1E1207.4-5209}

\author{Alexander~V.~Turbiner}
 \altaffiliation[]{On leave of absence from the Institute for
Theoretical and Experimental Physics, Moscow 117259, Russia}
\email{turbiner@nucleares.unam.mx}

\affiliation{%
Instituto de Ciencias Nucleares, Universidad Nacional Aut\'onoma
de M\'exico, Apartado Postal 70-543, 04510 M\'exico, D.F.,
M\'exico.}

\begin{abstract}
A model of a mixed helium-hydrogenic atmosphere of the isolated
neutron star 1E1207.4-5209 is proposed. It is based on the
assumption that the main components in the atmosphere are the
exotic molecular ions $He_2^{3+}$ and $H_3^{2+}$ with the presence
of $He^{+}, (HeH)^{2+}, H_2^{+}$ under a surface magnetic field
$\approx 4.4 \times 10^{13}$\,G. In addition to two absorption
features observed by {\it Chandra} observatory (Sanwal et al,
2002) the model predicts one more narrow absorption feature at
$\approx 400$\, eV.
\end{abstract}


\maketitle

Recently, it was discovered by Chandra $X$-ray observatory the
absorption features at 0.7\,KeV and 1.4\,KeV in the X-ray spectrum
of the isolated neutron star 1E1207.4-5209 \cite{Sanwal:2002} (see
Fig.1). Later this discovery was confirmed by XMM-Newton $X$-ray
observatory \cite{Bignami:2003}. These features had widths 0.1 and
0.13 KeV, respectively. For many people it was quite impressive
property that the second absorption energy is about twice larger
than the first one. Different proposals about content of the
atmosphere are presented. All of them are agreed on partially
ionized atmosphere content. Almost all of them are related to
atomic ions (for a review, see \cite{Sanwal:2002, Mori:2003}).
Among of these models it was proposed a model of the hydrogenic
atmosphere of 1E1207.4-5209 with main abundance of the exotic ion
$H_3^{2+}$ \cite{Turbiner:2004m}. In particular, this model
explains the observed absorption features as a consequence of the
photodissociation $H_3^{2+} \rar H + 2p$ and the $H_3^{2+}$
photoionization, while the bound-bound transitions are assumed to
be non-important. However, this model requires very strong
magnetic field on the surface, $(2-6) \times 10^{14}$\,G.

Above-mentioned discovery pushed us for searching new exotic
chemical compounds in a strong magnetic field in addition to ${\rm
H}_3^{2+}$ \cite{Turbiner:1999}. Very recently, we had discovered
that in the domain $ 4.4 \times 10^{13} \gtrsim B \gtrsim
10^{11}$\,G two more exotic systems can occur: the molecular ions
$({\rm HeH})^{2+}$ and ${\rm He}_2^{3+}$ \cite{Turbiner:2004}.
These exotic systems do not exist without magnetic field. Among
these two the one-electron molecular ion $He_2^{3+}$
\cite{Turbiner:2004} plays exceptionally important role at large
magnetic fields: at $B \gtrsim 3 \times 10^{12}$\,G it becomes the
Coulomb system with lowest total energy among one-electron systems
made from protons and/or $\alpha-$particles. It was also quite
surprising to find out that at the same domain of magnetic fields
another exotic ion $H_3^{2+}$ becomes the Coulomb system with
lowest total energy among one-electron systems made from protons.
It is worth emphasizing that in very large domain of magnetic
fields, $2 \times 10^{14} \gtrsim B \gtrsim 2 \times 10^{12}$\,G
the ratio of the binding (ionization) energies of these two
systems remains almost constant,
\[
   \frac{E_T^{H_3^{2+}}}{E_T^{He_2^{3+}}} \approx 0.5 \ .
\]

In this Note we propose a mixed helium-hydrogenic model of
atmosphere of the 1E1207.4-5209. This model provides an
explanation of both absorption features and requires a surface
magnetic field $\approx 4 \times 10^{13}$\,G, which is not in so
striking contradiction with that estimated using the period
derivative, $(2-4) \times 10^{12}$\,G (see \cite{Sanwal:2002}).

Let us make the simplest assumption that the atmosphere consists
of $\alpha-$ particles, protons and electrons mostly in a form of
the $H,\ H_2^+,\ H_3^{(2+)},\ H_4^{(3+)}, ({\rm HeH})^{2+}, He^+$
and ${\rm He}_2^{3+}$ systems \footnote{In principle, deuterons or
tritons can be instead of protons.}. The first important
observation is the fact that the only neutral system $H$ is always
characterized the highest total energy among above-mentioned
systems for any chosen magnetic field strength
\cite{Turbiner:1999,Turbiner:2004}. Hence, in general, it can be
neglected. It is evident that the charged systems $H_2^+,\
H_3^{(2+)},\ H_4^{(3+)}, ({\rm HeH})^{2+}, He^+$ and ${\rm
He}_2^{3+}$ can move mostly in the longitudinal (along the
magnetic line) direction, and their transverse motion is limited
to a domain defined by a Larmor radius. They are divided into
three types: ionization (bound-free transitions), dissociation and
excitation (bound-bound transitions). A study of all
above-mentioned systems were done using the variational technique
with physically relevant trial functions (see e.g.
\cite{Turbiner:1999, Turbiner:2004}. In Table I the binding
energies for magnetic field $4.4 \times 10^{13}$\,G are given.

It can be immediately seen that the binding (ionization) energies
of the most bound one-electron system $H_3^{2+}$ as well as
$H_2^{+}$ belong to the domain of the first absorption feature at
0.7 KeV. While the binding (ionization) energies of the $({\rm
HeH})^{2+}, He^+$ and ${\rm He}_2^{3+}$ systems correspond to the
second absorption feature at 1.4 KeV.
\begin{table*}[hp]
\label{Ionization}
    \caption{\it Binding energies in Rydbergs (Ry) and in
    electron-volts (eV) for one-electron Coulomb systems for magnetic
    field $4.4 \times 10^{13}$\,G. Energies in eV are rounded to the
    nearest integer number ending in 0 or 5.}
    \begin{ruledtabular}
\begin{tabular}{lcccccc}
\hline
 $H$-atom & $He^+$ & $He_2^{3+}$ & $(HeH)^{2+}$ &$H_2^{+}$ & $H_3^{2+}$ & \\
\hline
 33.8     & 93.5   & 105.1       & 92.6         & 54.5     &  55.2  & Ry\\
 460      & 1270   & 1430        & 1260         & 740      &  750   & eV\\
\hline
\end{tabular}
\end{ruledtabular}
\end{table*}
In Table II we present the dissociation energies. The range of
sensitivity of the {\it Chandra/ACIS} detector does not allow to
see the domain where the photodissociation process $He_2^{3+} \rar
He^+ + \alpha$ can contribute. In principle, two other
photodissociation processes $H_3^{2+} \rar H + 2p$ and $H_2^+ \rar
H + p$ can be seen although they are close to the end of the range
of sensitivity of the {\it Chandra/ACIS} detector.

\begin{table*}[htbp]
\label{Dissociation}
    \caption{\it Dissociation energies in Rydbergs (Ry) and
    electron-volts (eV) for some one-electron systems for
    magnetic field $4.4 \times 10^{13}$\,G. Energies in eV
    are rounded to the nearest integer number ending in 0 or 5.}
    \begin{ruledtabular}
\begin{tabular}{lccccc}
 \hline
 $He_2^{3+} \rar He^+ + \alpha$  & $H_2^+ \rar H + p$ & $H_3^{2+} \rar H + 2p$
 &  \\
\hline
 11.6 & 20.7 & 21.4 &  Ry\\
 160  & 280  & 290  &  eV\\
\hline
\end{tabular}
\end{ruledtabular}
\end{table*}

\begin{table*}[htbp]
\label{Excitation}
    \caption{\it Excitation energies in Rydbergs (Ry) and
   electron-volts (eV) for some one-electron systems for
   magnetic field $4.4 \times 10^{13}$\,G. Energies in eV
   are rounded to the nearest integer number ending in 0 or 5.}
    \begin{ruledtabular}
\begin{tabular}{lccccc}
 \hline
 & $He_2^{3+} (1\sigma_g \rar 1\pi_u) $
 & $H_2^{+}   (1\sigma_g \rar 1\pi_u) $
 & $H_3^{2+}  (1\sigma_g \rar 1\pi_u) $
 &  &  \\
\hline
  & 29.0  & 13.4 & 15.0 &  & Ry\\
  & 395   & 180  & 205  &  & eV\\
\hline
\end{tabular}
\end{ruledtabular}
\end{table*}

In Table III we present the energies of the first excitation of
$H_2^{+}$ and $H_3^{2+}$, correspondingly. These processes
contribute to the domain which is close to the end of the range of
sensitivity of the {\it Chandra/ACIS} detector and unlikely to be
detected. The first excitation of $He_2^{3+}$ can be seen. This
line can be quite narrow. Hence, the domain around 0.4 KeV
deserves to be observed carefully. We assume that the
photoexcitation cross-sections are smaller than the
photoionization ones following the results obtained for the
hydrogen atom \cite{Potekhin:2003}.

%\newpage
\begin{figure}
\begin{center}
   \psfig{file=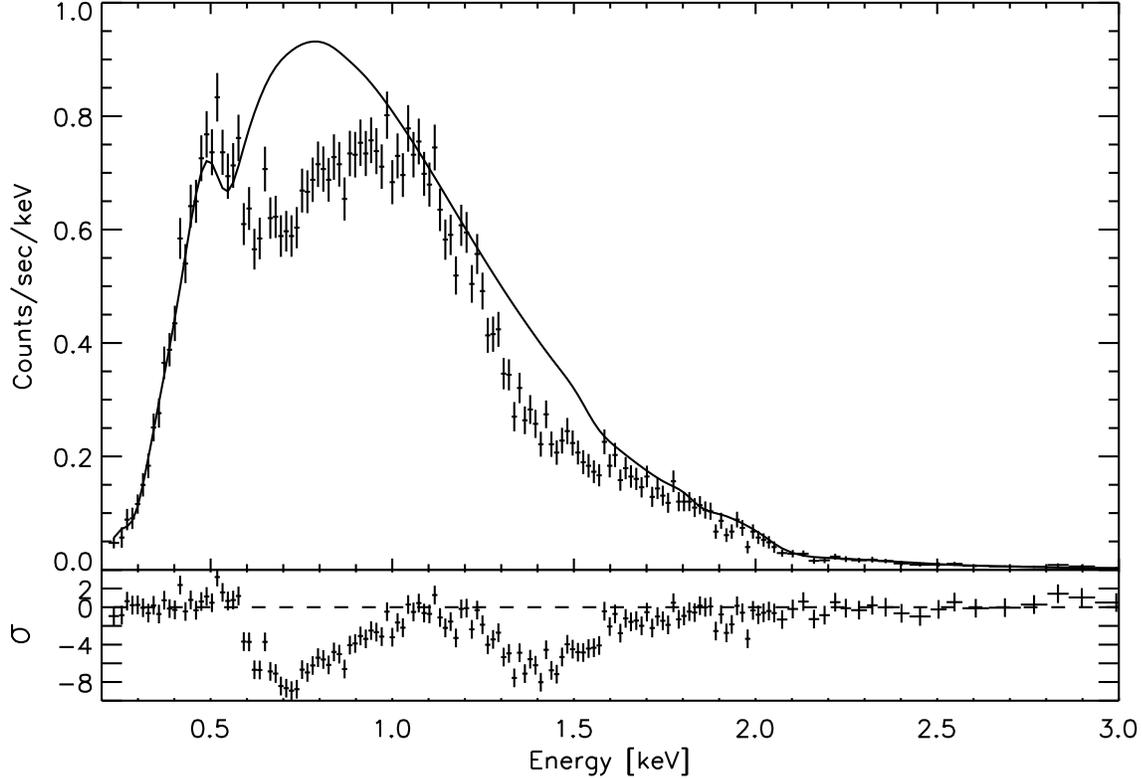,width=150mm,angle=0}
     \caption{{\it Chandra}/ACIS spectrum as presented
     in \cite{Mori:2003} (Fig.\#1 from \cite{Mori:2003} by the
     author's permission).
     The solid line is a blackbody model of \cite{Mori:2003}
     which is corrected by the detector sensitivity to illustrate
     two absorption features.}
\end{center}
\end{figure}

We can formulate our model of the content of the neutron star
atmosphere. It corresponds to the surface magnetic field $\approx
4.4 \times 10^{14}$\,G. The atmosphere is partially ionized and
consists of $H_2^+,\ H_3^{(2+)},\ He^+,\ ({\rm HeH})^{2+},\ {\rm
He}_2^{3+}$ ions. Photoionization of $H_2^+, H_3^{2+}$ explains
the first absorption feature in Fig.1 at 0.7 KeV while
photoionization $He^+,\ ({\rm HeH})^{2+},\ {\rm He}_2^{3+}$
explains the second absorption feature at 1.4 KeV
 \footnote{The cross-section of photo-ionization
depends on the energy of ionization. We unaware of any reliable
calculations of bound-bound and bound-free transitions, even for
the simplest molecular system $H_2^+$, we assume, following a
detailed study of the hydrogen atom \cite{Potekhin:2003}, that (i)
photo-ionization cross-section has a maximum near the ionization
threshold, and (ii) bound-bound transition amplitudes are small
compared to the bound-free ones. We also neglect any difference
between ionization threshold (binding energy) and the maximum of
the energy distribution, assuming that this difference is small.}.
We predict a narrow absorption feature emerging from electronic
excitation of $He_2^{3+} (1\sigma_g \rar 1\pi_u) $ around 400 KeV.
It can allow to identify the present model. Red-shift effects are
not taken into account, but they can be included straightforward
(see e.g. \cite{Sanwal:2002}) and we guess can increase the
magnetic field values for $\sim 20-50\%$. We think that the
presence of the gravitational sedimentation of hydrogen and helium
will not change our results.

The model of the atmosphere with main abundance of the $He^+$
atomic ions proposed in \cite{Sanwal:2002} with the surface
magnetic field $1.5 \times 10^{14}$\,G is lacking the explanation
why a possible presence of the molecular ion $He_2^{3+}$ is
neglected. For such a magnetic field the molecular ion $He_2^{3+}$
is much more bound than the atomic ion $He^+$. It can be
calculated \footnote{In non-relativistic framework, neglecting the
center-of-mass effects and relativistic corrections, which we
assume following the Salpeter et al (see \cite{Lai:2001}) do not
exceed 10\% in the binding energy.} that total energy of
$He_2^{3+}$ at $B=1.5 \times 10^{14}$\,G is lower than the total
energy of $He^+$ to $\approx 26.4$\,Ry. Even under NS surface
temperature 1.4-1.9 MK (see, e.g. \cite{Sanwal:2002}) it is not
clear why the presence of $He_2^{3+}$ should be discarded.
Perhaps, it is worth mentioning that for the magnetic field $1.5
\times 10^{14}$\,G another exotic molecular ion $(HeH)^{2+}$ is
also more bound than $He^+$:
\[
E_T^{He^+} - E_T^{(HeH)^{2+}} \approx 4.4 \, \mbox{Ry}\ .
\]

The author is grateful to J.C.~Lopez Vieyra. K.~Mori, G.G.~Pavlov
and A.Y.~Potekhin for valuable discussions. The work is supported
in part by CONACyT grant {\it 47899-E}.

\end{document}